\input jnl
\input reforder
\input eqnorder
\input epsf.tex

\def\3he{{$^3${\rm He}}}

\def\eg{{\it e.g.,\ }}
\def\ie{{\it i.e.,\ }}
\def\etal{{\it et al.}}
\def\etc{{\it etc.}}

\def\slD{\raise.15ex\hbox{$/$}\kern-.53em\hbox{$D$}}
\def\dsl{\raise.15ex\hbox{$/$}\kern-.57em\hbox{$\Delta$}}
\def\slp{{\raise.15ex\hbox{$/$}\kern-.57em\hbox{$\partial$}}}
\def\nsl{\raise.15ex\hbox{$/$}\kern-.57em\hbox{$\nabla$}}
\def\sla{\raise.15ex\hbox{$/$}\kern-.57em\hbox{$\rightarrow$}}
\def\slla{\raise.15ex\hbox{$/$}\kern-.57em\hbox{$\lambda$}}
\def\slb{\raise.15ex\hbox{$/$}\kern-.57em\hbox{$b$}}
\def\lnp{\raise.15ex\hbox{$/$}\kern-.57em\hbox{$p$}}
\def\lnk{\raise.15ex\hbox{$/$}\kern-.57em\hbox{$k$}}
\def\lnK{\raise.15ex\hbox{$/$}\kern-.57em\hbox{$K$}}
\def\lnq{\raise.15ex\hbox{$/$}\kern-.57em\hbox{$q$}}

\def\psib{{\overline\psi}}

\def\cL{{\cal L}}
\def\cM{{\cal M}}

\def\cO{{\cal O}}


\def\pmb#1{\setbox0=\hbox{$#1$}%
\kern-.025em\copy0\kern-\wd0
\kern.05em\copy0\kern-\wd0
\kern-.025em\raise.0433em\box0 }

\def\q2{{Q^2}}
\def\gtwid{\raise.3ex\hbox{$>$\kern-.75em\lower1ex\hbox{$\sim$}}}
\def\ltwid{\raise.3ex\hbox{$<$\kern-.75em\lower1ex\hbox{$\sim$}}}
\def\12{{1\over2}}
\def\part{\partial}

\def\low#1{\lower.5ex\hbox{${}_#1$}}

\def\psl{\raise.15ex\hbox{$/$}\kern-.57em\hbox{$\partial$}}
\def\partt{\raise.15ex\hbox{$\widetilde$}{\kern-.37em\hbox{$\partial$}}}

\def\topppageno1{\global\footline={\hfil}\global\headline
={\ifnum\pageno<\firstpageno{\hfil}\else{\hss\twelverm --\ \folio
\ --\hss}\fi}}
 
\def\toppageno2{\global\footline={\hfil}\global\headline
={\ifnum\pageno<\firstpageno{\hfil}\else{\rightline{\hfill\hfill
\twelverm \ \folio
\ \hss}}\fi}}

\def\prl#1{Phys.\ Rev.\ Lett.\ {\bf #1}}
\def\prd#1{Phys.\ Rev.\ {\bf D#1}}
\def\plb#1{Phys.\ Lett.\ {\bf #1B}}
\def\npb#1{Nucl.\ Phys.\ {\bf B#1}}

\def\ie{{\it i.e.},\ }
\def\eg{{\it e.g.},\ }

\def\ref#1{${}^{#1}$}
\def\nsection#1 #2{\leftline{\rlap{#1}\indent\relax #2}}

\def\prl#1{Phys.\ Rev.\ Lett.\ {\bf #1}}
\def\prd#1{Phys.\ Rev.\ {\bf D#1}}
\def\plb#1{Phys.\ Lett.\ {\bf #1B}}
\def\npb#1{Nucl.\ Phys.\ {\bf B#1}}

\def\qa{{quenched approximation}}
\def\chpt{{chiral perturbation theory}}
\def\psib{{\overline{\psi}}}

\def\etat{{\tilde{\eta}}}
\def\Dsl{D\!\!\!\!/\,\,}
\def\eq#1{{eq. \(#1)}}
\def\eqs#1{{eqs. \(#1)}}

\def\etc{{\it etc}}
\def\ie{{\it i.e.}}
\def\eg{{\it e.g.}}
\def\qt{{\tilde q}}
\def\qbar{{\overline q}}
\def\qtbar{{\overline\qt}}

\def\dbar{{\overline d}}
\def\sbar{{\overline s}}
\def\phit{{\tilde\phi}}
\def\half{{{1\over 2}}}
\def\cf{{\it cf.}}
\def\mbar{{\overline m}} 
\def\Kbar{{\overline K}} 
\def\ZZ{{Z\!\!\!Z}}

{\parindent=0pt {October 1994 \hfill{Wash. U. HEP/94-63} }
\rightline{hep-lat/9411005}

\title Chiral Perturbation Theory and the Quenched Approximation of QCD$\null^*$
\footnote{}{${}^*$ Lectures given at the XXXIV Cracow School of Theoretical\ %
Physics, Zakopane, Poland}
\author Maarten F.L. Golterman$\null^{**}$
\footnote{}{${}^{**}$ e-mail: maarten@aapje.wustl.edu}                       
\affil Department of Physics 
       Washington University 
       St. Louis, MO 63130, USA 

\abstract {The \qa\ for QCD is, at present and in the foreseeable
future, unavoidable in lattice calculations with realistic choices of
the lattice spacing, volume and quark masses. In these lectures, 
I review the
analytic study of the effects of quenching based on \chpt. 
Quenched \chpt\ leads to quantitative
insight into the difference between quenched and unquenched QCD, and
reveals clearly diseases which plague
quenched QCD.  A short review of the ideas underlying
\chpt\ is included.} 
\endtopmatter

\subhead{\bf 1. Introduction}

The lattice formulation of QCD has proven to be a powerful tool for
computing QCD quantities of direct phenomenological interest, such as
hadron masses, decay constants, weak matrix elements, the strong
coupling constant, \etc. (For reviews see for instance refs.
[\cite{sharpe,mackenzie}], the proceedings of Lattice 93
[\cite{lat93}], and the lectures of Rajan Gupta in these proceedings.)
 
In order to perform such computations numerically, one obviously needs 
to consider a system with a finite number of degrees of freedom, which 
is accomplished by putting lattice QCD in a finite box.  This box is 
then hopefully large enough to accomodate the physics one is interested 
in without serious finite size effects.  This leads to the (minimal)
requirement 
that the Compton wavelength of the particles of interest is sufficiently 
smaller than the linear dimension of the box, \ie\ the mass has to be 
large enough for the particles to fit in the box. 

In order to have a small enough lattice spacing, small enough masses (in
particular for the pion) and a large enough box size, one needs a large
number of degrees of freedom in a numerical computation. It turns out
that for QCD with realistic choices of the lattice spacing $a$, volume
$V$ and the quark masses (in particular the light quark masses), the
presently available computational power is not adequate. The most severe
problem comes from the fermion determinant, the logarithm of which is a
very nonlocal part of the gluon effective action (specially for light
quark masses). This nonlocality slows down currently available algorithms
dramatically.

In order to ``circumvent" this problem, most numerical computations in
lattice QCD have been done in the \qa, in which one simply replaces the
fermion determinant by one [\cite{quenched}]. This amounts to
ignoring all fermion loops which occur in QCD correlation functions
(except those put in by hand through the choice of operators on the
external lines). While some handwaving arguments exist as to why this
might not be unreasonable, the \qa\ does introduce an uncontrolled
systematic error. Since the effect of a fermion loop is roughly
inversely proportional to the fermion mass, this error is expected to be
particularly large for quantities involving light quarks. 
Therefore one might expect
that \chpt\ (ChPT) is a useful tool for
investigating the difference between quenched and unquenched (``full")
QCD.

In this talk, I will review a systematic approach to the study of the
\qa\ through ChPT [\cite{sharpe1,us1, sharpe2,uslat92}]. 
There are two
reasons why ChPT is useful in this context:

$\bullet$ It turns out that ChPT can be systematically adapted to
describe the low energy sector of quenched QCD [\cite{us1}]. It will
therefore give us nontrivial, quantitative information on the difference
between quenched and full QCD.

$\bullet$ ChPT describes the approach to the chiral limit, and can be
used for extrapolation of numerical results to small masses and large
volumes. If these results come from quenched computations, one will of
course need a quenched version of ChPT. (For finite volume ChPT, see
refs. [\cite{hasleu}]. For quenched finite volume results, see refs.
[\cite{us1,sharpe2}].)

In this review, I will concentrate on the first point. 
First, I will give a short summary of the basic underpinnings of ChPT, 
in order to make these lectures more or less selfcontained.  I will then
give an early example of numerical results which can be understood using
one loop ChPT.  I will then go on to show
how ChPT is developed for the \qa, and use it for a quantitative
comparison between full and quenched QCD. The quantities that I will
discuss are $f_K/f_\pi$ [\cite{us1,uslat92,rajan94}] 
and the octet baryon masses
[\cite{labsha}].

I will then address a number of theoretical problems that arise as a 
consequence of quenching.  That such problems arise is no surprise, 
as quenching QCD mutilates the theory quite severely.  It is however 
quite instructive to see what the actual consequences are.  

\subhead{\bf 2. A review of ChPT}

In this section I will summarize the basic ideas of ChPT, which is an 
efficient way to organize the information we can obtain from the QCD 
Ward identities for chiral symmetry.  A partial 
list of useful references is [\cite{weinberg,georgi,gasleu}]. 
Let us start with QCD with three flavors.  The fermion part of the 
lagrangian is
$$\cL_{\rm QCD}=\qbar_L\Dsl q_L+\qbar_R\Dsl q_R+\qbar_LMq_R+\qbar_R
M^\dagger q_L.\eqno(qcdl)$$
$q=(u,d,s)$ is a three flavor quark field, and the subscripts $L$ and 
$R$ denote the left- and righthanded projections $q_{L,
R}=\half(1\pm\gamma_5)q$.  $M$ is the quark mass 
matrix
$$M=\pmatrix{m_u & 0 & 0 \cr 0 & m_d & 0 \cr 0 & 0 & m_s 
\cr}.\eqno(mass)$$
For $M=0$, the lagrangian is invariant under the chiral symmetry group
$U(3)_L\times U(3)_R$:
$$q_L\to U_Lq_L,\ \ \ q_R\to U_Rq_R,\ \ \ U_L,U_R\in U(3).\eqno(trans)$$
Note that formally \eq{qcdl}\ is invariant if we also let $M$ transform 
as
$$M\to U_LMU_R^\dagger.\eqno(Mtrans)$$
Axial $U(1)$ transformations 
(\ie\ $U(1)$ transformations on the fields $q_L$ and $q_R$ for which the 
phases are not equal) however are not a symmetry of the theory, since 
they are broken by the axial anomaly. The real symmetry group is therefore
$SU(3)_L\times SU(3)_R\times U(1)$ where the $U(1)$ just corresponds
to quark number conservation.  It is furthermore believed that (still 
for $M=0$) this symmetry is spontaneously broken to $SU(3)_V\times 
U(1)$, where $SU(3)_V$ is the group of transformations for which $U_L=U_R\in 
SU(3)$.  The ensuing eight Goldstone bosons are the mesons $\pi^\pm$, $\pi^0$,
$K^\pm$, $K^0$, $\Kbar^0$ and $\eta$, which transform in the 
octet representation of $SU(3)$, denoted as the hermitian, traceless
$3\times 3$ matrix $\phi$:
$$
\phi=\pmatrix{{{\pi^0}\over{\sqrt{2}}}+{{\eta}\over{\sqrt{6}}} & \pi^+ & 
K^+ \cr \pi^- & -{{\pi^0}\over{\sqrt{2}}}+{{\eta}\over{\sqrt{6}}} & K^0 
\cr K^- & \Kbar^0 & -{{2\eta}\over{\sqrt{6}}} \cr}.
\eqno(fulloctet)$$
These mesons acquire a nonzero mass due to the quark mass matrix $M$, as 
we will discuss below.

One can now try to describe the physics of hadrons by constructing an 
effective lagrangian for composite operators with the quantum numbers of 
these hadrons.  Explicitly, the operator $H_{ij}=q_{Li}\qbar_{Rj}$ 
(summed over color and spin indices, $i$ and $j$ are flavor indices) 
will couple to the Goldstone mesons, and one can introduce other 
operators for other hadrons, and write down an effective lagrangian of 
the form
$$\cL_{\rm eff}(H, H^\dagger, {\rm baryons}, ...).\eqno(leff)$$
This lagrangian, integrated over space-time, should be invariant under
$SU(3)_L\times SU(3)_R\times U(1)$ transformations $H\to 
U_LHU_R^\dagger$ \etc.
We can decompose $H=R\Sigma$ with $R$ hermitian and positive, and 
$\Sigma$ unitary.  
The spontaneous symmetry breakdown $SU(3)_L\times 
SU(3)_R\times U(1)\to SU(3)_V\times U(1)$ corresponds to $R$ acquiring a 
nonzero vacuum expectation value $r{\bf 1}$.  Dropping the other fields 
in $\cL_{\rm eff}$, the effective potential does not 
depend on $\Sigma$, and
$\Sigma=exp(2i\phi/f)$ can therefore be 
identified with the Goldstone bosons, \eq{fulloctet}, leading to the nonlinear 
realization of chiral symmetry [\cite{oldies}].  $f$ is a constant with 
the dimension of a mass.  (The overall phase of 
$\Sigma$ corresponds to the $\eta'$ which is not a Goldstone boson due 
to the anomaly [\cite{thooft,witven}], and has a mass of order $1$ $GeV$.)  
The fluctuations of $\phi$ describe 
the Goldstone mesons, whereas fluctuations in $R$ around $r{\bf 1}$ describe 
other heavy (scalar) mesons (which may or may not exist as narrow 
resonances in nature).

At low energies (energies below the masses of any non-Goldstone hadrons) 
the effective lagrangian simplifies to
$$\cL_{\rm eff}(\Sigma,M),\eqno(leffg)$$
which has to be invariant under the transformations
$$\eqalignno{
\Sigma&\to U_L\Sigma U_R^\dagger,\cr
M&\to U_LMU_R^\dagger.
&(leffsymm)}$$
This follows from the definition of $H$, which, ignoring fluctuations in 
$R$, is $H=r\Sigma$ and from the invariance of the QCD path 
integral under the transformations \eqs{trans,Mtrans}.  Note that the 
unbroken symmetry $SU(3)_V$ is linearly realized ($\Sigma\to 
V\Sigma V^\dagger$ implies $\phi\to V\phi V^\dagger$), whereas the 
broken symmetries are nonlinearly realized.

Let us imagine that we can expand $\cL_{\rm eff}$ in terms of 
derivatives and the quark mass matrix $M$.  To second order in 
derivatives and linear order in $M$, (we will call this ``$O(p^2)$")
the most general expression is\footnote{${}^1$}{\tenrm
We will not need to consider
the Wess--Zumino term, see ref. [\cite{gasleu}] and refs. therein.} 
$$\cL_{\rm eff}={{f^2}\over 8}\;tr(\partial_\mu\Sigma\partial_\mu
\Sigma^\dagger)-v\;tr(M\Sigma^\dagger+M^\dagger\Sigma),\eqno(leffsnd)$$
where $f$ and $v$ are undetermined constants.  Expanding $\cL_{\rm eff}$
to quadratic order in the meson field $\phi$ (\eq{fulloctet})
using $\Sigma=1+{{2i\phi}\over f}+...$ one will see that the kinetic 
terms have the standard normalization, and one can read off the tree 
level meson 
masses (for simplicity I will choose $m_u=m_d=m$):
$$m_\pi^2={{8vm}\over{f^2}},\ \ \ m_K^2={{4v(m+m_s)}\over{f^2}},
\ \ \ m_\eta^2={{8v(m+2m_s)}\over{3f^2}}.\eqno(mesonmasses)$$
We see that the mesons acquire masses proportional to the square root of 
the quark masses, and in the chiral limit $M\to 0$ they are massless as 
they should be.  Note that for onshell mesons the combined expansion to 
second order in derivatives and to first order in $M$ is consistent.  
Note also that we obtained our first nontrivial result: from 
\eq{mesonmasses}\ it follows that
$$m_\eta^2={4\over 3}m_K^2-{1\over 3}m_\pi^2,\eqno(meta)$$
which predicts a value of $m_\eta$ about $3\%$ too large.

Let us proceed, and calculate scattering amplitudes $A$ to some 
order in the loop expansion.  In order to do this we need to introduce a 
cutoff $\Lambda$.  The physical reason for this is that we are now 
ignoring all the hadronic physics at energies of order the $\rho$ mass 
and beyond.  We therefore expect that we will have to choose $\Lambda$ 
to be of the order of the $\rho$ mass, \ie\ or order $1\;GeV$.
Using dimensional regularization and  
power counting one obtains for the contribution of a certain diagram to 
$A$ [\cite{weinberg,georgi}]
$$(2\pi)^4\delta(\sum_i p_i)f^2p^2\left({{p^2}\over{(4\pi 
f)^2}}\right)^N\left({1\over f}\right)^E F(p^2/\Lambda^2).\eqno(ampl)$$
We identified $\Lambda\approx 4\pi f$, which is of order $1\;GeV$ 
($f$ will turn out to be the pion decay constant, as we will see in a 
moment)
[\cite{georgi}].
Here $p_i$ stands for external momenta, and $p^2$ denotes the square of any 
linear combination of these; it
can also be a 
Goldstone meson mass squared.  $E$ is the number of external lines, and
$$N=\sum_d\half(d-2)V_d+L,\eqno(N)$$
where $L$ is the number of loops and $V_d$ is the number of vertices 
with $d$ derivatives (or $d/2$ powers of $M$, or any combination in 
between).  $F$ is a function containing logarithmic divergences in 
$\Lambda$ arising from diagrams with loops.

From \eq{ampl}\ we can draw the following conclusions.  An amplitude $A$ 
can be calculated as a series expansion in $p^2/(4\pi f)^2$ if ${\vec 
p}^2$ and $m_{\rm meson}^2$ are much smaller than $\Lambda^2=(4\pi 
f)^2$.  In \eq{leffsnd}\ we only allowed $O(p^2)$ terms, but we could 
have allowed higher order terms.  From \eq{ampl}\ we see that treelevel 
contributions from $O(p^4)$ terms come in at the same order as one loop 
terms from the $O(p^2)$ lagrangian \eq{leffsnd}: $N=1$ corresponds to 
$L=1$ and $V_4=0$ or $L=0$ and $V_4=1$.  Therefore the $O(p^4)$ terms 
act as counterterms at the one loop level, and in fact need to be added 
at this order in order to absorb the cutoff dependence (\cf\ $F$ in 
\eq{ampl}).  A change in the choice of $\Lambda$ can be absorbed by a 
shift in these counterterms.
These conclusions generalize systematically to higher loops 
($L>1$).  A very important conclusion of this analysis is that at any 
given order in $p^2/(4\pi f)^2$ we only need a finite number of 
counterterms, and therefore the theory is predictive.  Furthermore, the 
nonanalytic terms at $L$ loops (contained in $F$) are determined by the 
$O(p^{d=2L})$ lagrangian (\ie\ by diagrams with $V_d=0$ for $d>2L$).

As an example I will now discuss the decay constants $f_\pi$ and $f_K$.
First consider $N=0$, so that the $O(p^2)$ lagrangian at tree level is 
sufficient.  The pion decay constant $f_\pi$ in QCD is defined by the matrix 
element of the lefthanded Noether
current $j_{L\mu}=\dbar_L\gamma_\mu u_L$ 
between a one pion state and the vacuum:
$$\langle 0|j_{L\mu}(x)|\pi^+(p)\rangle={i\over 2}p_\mu f_\pi e^{-ipx}.
\eqno(fpidef)$$
(It is this current which couples to the electroweak $W$ bosons, through 
which the pion decays into a pair of leptons.)
In ChPT this Noether current can be determined from \eq{leffsnd}, 
leading to
$$j_{L\mu}={{if^2}\over 4}\;tr(T^+\partial_\mu\Sigma\Sigma^\dagger)
=-{f\over 2}\partial_\mu\pi^++...,
\eqno(current)$$
where $T^+$ is the appropriate $SU(3)$ generator, and hence
$$\langle 0|j_{L\mu}(x)|\pi^+(p)\rangle={i\over 2}p_\mu f e^{-ipx}.
\eqno(fpichpt)$$
We conclude that at this order $f_\pi=f\;(=132\;GeV)$ which determines the 
constant $f$.  A similar calculation leads to 
$$f_K=f=f_\pi,\eqno(fk)$$
which is a prediction, off by $22\%$.

We see that the relation between $f_K$ and $f_\pi$ is consistent with 
$SU(3)_V$ symmetry.  The $SU(3)_V$ breaking introduced by the quark 
masses shows up at the one loop order ($N=1$).  From ref. 
[\cite{gasleu}]
$${{f_K}\over{f_\pi}}=1+{5\over 4}{{m_\pi^2}\over{16\pi^2f_\pi^2}}
\log{\Bigl({{m_\pi^2}\over{\Lambda^2}}\Bigr)}
-{1\over 2}{{m_K^2}\over{16\pi^2f_\pi^2}}
\log{\Bigl({{m_K^2}\over{\Lambda^2}}\Bigr)}
-{3\over 4}{{m_\eta^2}\over{16\pi^2f_\pi^2}}
\log{\Bigl({{m_\eta^2}\over{\Lambda^2}}\Bigr)}
+{{m_K^2-m_\pi^2}\over{16\pi^2f_\pi^2}}L,\eqno(fullfkfpi)$$
where $L$ is a linear combination of coefficients of $O(p^4)$ terms in 
the effective lagrangian [\cite{gasleu}].  With $\Lambda=1\;GeV$ the 
experimental value $f_K/f_\pi=1.22$ can be reproduced with $L=1.1\/$.
Notice that the coefficient 
of $\log{(1/\Lambda^2)}$ is proportional to $m_K^2-m_\pi^2$, 
so a change in the 
cutoff can be absorbed by a shift in $L$.  We also see that the 
expansion parameters are
$${{m_\pi^2}\over{16\pi^2f_\pi^2}}\approx 0.007,
\ \ \ {{m_K^2}\over{16\pi^2f_\pi^2}}\approx 0.09,\eqno(exppar)$$
which are small.  In contrast, 
${{m_D^2}\over{16\pi^2f_\pi^2}}\approx 
1.3$, and hence the charm quark mass cannot be considered as small in 
the sense of \chpt.  

An equation like \eq{fullfkfpi}\ can be used in two ways.  First, we can 
determine $L$ from the experimental values for $f_K$ and $f_\pi$ (fixing 
$\Lambda$ at some value of order $1\;GeV$).  This can then be used to 
predict other quantities, like $f_\eta$ [\cite{gasleu}].  Second, 
\eq{fullfkfpi}\ can be used to fit results from numerical computations.  
Since these usually are performed at relatively large values of the 
quark mass (and hence the Goldstone meson masses), we can then employ 
results from ChPT to extrapolate to the physical values of quark masses.

\subhead{\bf 3. An example: $B_K$}

In this section, I would like to give an example of the 
definition and use 
of an electroweak operator in ChPT (for much more detail see
\eg\ refs. [\cite{georgi,cbtasi}] and refs. therein). 
The kaon $B$ parameter, $B_K$, which determines the strength of 
$K^0-\Kbar^0$ mixing, is defined as
$${8\over 3}f_K^2m_K^2B_K=\langle\Kbar^0|\cO_K|K^0\rangle,
\eqno(defB)$$
with
$$\cO_K=(\sbar_L\gamma_\mu d_L)(\sbar_L\gamma_\mu d_L).
\eqno(dstwo)$$
$\cO_K$ transforms as a component of the $(27,1)$ 
representation of $SU(3)_L\times SU(3)_R$.  This symmetry property can 
be used to construct a corresponding operator in ChPT, which to $O(p^2)$ 
is uniquely given by
$$\cO_K^{\rm ChPT}={1\over 3}Bf^4(\partial_\mu\Sigma\Sigma^\dagger)_{ds}
(\partial_\mu\Sigma\Sigma^\dagger)_{ds}.\eqno(dschpt)$$
(Note that in both QCD and ChPT this operator is the product of two 
lefthanded currents.)  The coefficient $B$ is undetermined, and is therefore 
another free parameter in ChPT, like $f$ and $v$.  To one loop order, we may 
calculate $B_K$ from \eqs{defB,dschpt}, and we obtain (for degenerate 
quark masses, \cf\ ref. [\cite{bijsonwise,sharpe2}] 
for the nondegenerate case)
$$B_K=B\left(1-6{{m_\pi^2}\over{16\pi^2f_\pi^2}}
\log{{{m_\pi^2}\over{\Lambda^2}}}
\right)+O(p^4)\ {\rm contributions}.
\eqno(bkfull)$$

In ref. [\cite{sharpe2}], besides $B_K$, 
a different quantity $B_V$ was 
studied, which, as we will see, has chiral one loop corrections larger 
than $B_K$ has.  For all details not discussed here I refer to ref. 
[\cite{sharpe2}].  The quantity $B_V$ is defined from
$${4\over 3}f_K^2m_K^2B_V=
\langle{\Kbar'}{}^0|[(\sbar'_a\gamma_\mu d'_b)(\sbar_b\gamma_\mu d_a)
+(\sbar'_a\gamma_\mu d'_a)(\sbar_b\gamma_\mu 
d_b)]|K^0\rangle,\eqno(bv)$$
where $a$ and $b$ are color indices.  $d'$ and $s'$ are new flavors of 
quarks, and ${\Kbar'}{}^0$ is a kaon built out of those.  This is a 
technical trick to reduce the number of Wick contractions on \eq{bv}\ to 
one.  All quark masses are chosen equal.  Note that I use a 
normalization for the meson decay constants which is different from that 
used in ref. [\cite{sharpe2}].  With $B_A$ defined similarly 
with $\gamma_\mu\to\gamma_\mu\gamma_5$ we have
$$B_K=B_V+B_A.\eqno(kva)$$
One now can calculate $B_V$ to one loop in ChPT, where in this special 
case (with degenerate quark masses)
one can argue that the result is the same in the quenched and 
unquenched theories.  The result is
$$B_V=\half B_K-{3\over 4}\beta{{v^2}\over{m_K^2}}I_2(m_K^2)
-{3\over 8}\delta-{3\over 8}\gamma I_2(m_K^2)+O(m_K^2),
\eqno(bvoneloop)$$
where
$$I_2(m^2)={{m^2}\over{f^2}}\int{{d^4p}\over{(2\pi)^4}}
{1\over{(p^2+m^2)^2}}.\eqno(int)$$
$\beta$, $\gamma$ and $\delta$ are coefficients of operators in ChPT 
which arise in the calculation of $B_V$ to one loop
(contrary to the case of $B_K$ 
there is more than one [\cite{sharpe2}]).  $B_A$ is given by the same 
expression with opposite signs for $\beta$, $\gamma$ and $\delta$.

$$
\epsfbox{fig1.ps}
$$\nobreak
\centerline{
Figure 1. {\it Results for $B_V$ versus $m_\pi^2\;(=m_K^2)$ in lattice units
(from ref. [\cite{sharpe2}]).}} \medskip

From \eq{bvoneloop}\ we see why it is interesting to consider $B_V$.  
Unlike in $B_K$ there appears an ``enhanced" chiral logarithm in $B_V$
[\cite{langpag}] 
--- the second term on the righthand side of \eq{bvoneloop}.  The extra 
factor $m_K^2$ in the denominator, if small enough, will enhance the 
size of this correction.  If we repeat the calculation in a finite 
volume $V=L^3$, $B_V$ becomes $L$ dependent because the spatial part of 
the integral $I_2$ 
now has to be replaced by a one loop momentum sum over momenta
${\vec p}=2\pi{\vec n}/L$ with ${\vec n}\in\ZZ^3$
(with periodic boundary conditions).  Volume 
dependence of this nature has actually been seen in numerical results.
Fig. 1 is from ref. [\cite{sharpe2}].  In this graph the points 
represent quenched 
numerical results.  The solid lines show the result of a fit 
of the parameters $\half B_K-{3\over 8}\delta$, $\beta$, and 
$\gamma$ (for a precise explanation of 
the fit see again ref. [\cite{sharpe2}]).  The graph shows that the 
numerical results are within errors consistent with one loop ChPT.
Because the enhanced logarithms do not appear in $B_K$ itself, one loop 
effects in this case are too small to be
seen with the current statistical errors present
in the numerical results for $B_K$.

This concludes our first example of a confrontation of one loop ChPT with 
numerical computations.  In this special case, the results from ChPT are 
unchanged by the effects of quenching.  This is in general not true, and 
a systematic way of changing 
ChPT to correspond to the \qa\ needs to be developed. 
I will do this in the next section, and then return to other, more 
recent, examples.

\subhead{\bf 4. Systematic ChPT for quenched QCD}

In this section I will outline the construction of a chiral effective
action for the Goldstone boson sector of quenched QCD [\cite{us1}]. I
will first introduce the formalism, and then show how it works in some
examples. For early ideas on quenched ChPT, see ref. 
[\cite{morel,sharpe1}].

We will start from a lagrangian definition of euclidean quenched QCD. 
(We will restrict ourselves entirely to the euclidean theory which can 
be defined by a pathintegral.  Hamiltonian quenched QCD presumably does 
not exist.)  To
the usual QCD lagrangian with three flavors of quarks $q_a$, $a=u,d,s$,
we add three ghost quarks $\qt_a$ with exactly the same quantum numbers
and masses $m_a$, but with opposite, bosonic, statistics [\cite{morel}]: 
$$\cL_{\rm quarks}=
\sum_a\qbar_a(\Dsl+m_a)q_a+\sum_a\qtbar_a(\Dsl+m_a)\qt_a,\eqno(qqcd)$$ 
where $\Dsl$ is the covariant derivative coupling the quarks and ghost
quarks to the gluon field. The gluon effective action produced by
integrating over the quark- and ghost quarkfields vanishes, since the
fermion determinant of the quark sector is exactly cancelled by that of
the ghost sector.  Note that the ghost quarks violate the 
spin-statistics theorem.  Eq. \(qqcd) is the lagrangian definition of 
quenched QCD.

We will now assume that mesons are formed as (ghost) quark - (ghost)
antiquark pairs just like in ordinary QCD. This is basically equivalent
to the notion that it is the dynamics of the gluons which leads to
confinement and chiral symmetry breaking. The Goldstone particle
spectrum of quenched QCD will then contain not only $q\qbar$, but also
$\qt\qtbar$, $q\qtbar$ and $\qt\qbar$ bound states.  We will denote this 
36-plet by
$$\Phi\equiv\left(\matrix{\phi&\chi^\dagger\cr
\chi&\phit\cr}\right)\sim
\left(\matrix{q\qbar&q\qtbar\cr\qt\qbar&\qt\qtbar\cr}\right).\eqno(octet)$$
Note that the fields $\chi$ and $\chi^\dagger$ describe Goldstone 
fermions.

The quenched QCD lagrangian \eq{qqcd} with zero quark masses has a
much larger symmetry group than the usual $U(3)_L\times U(3)_R$ chiral
group; it is invariant under the graded group $U(3|3)_L\times U(3|3)_R$
[\cite{us1}],
where $U(3|3)$ is a graded version of $U(6)$ since it mixes the fermion
and boson fields $q$ and $\qt$. Writing an element $U$ of $U(3|3)$ in
block form as
$$U=\left(\matrix{A&C\cr D&B\cr}\right),\eqno(element)$$
the $3\times 3$ matrices $A$ and $B$ consist of commuting numbers, while 
the $3\times 3$ matrices $C$ and $D$ consist of anticommuting numbers.

We can now construct a low energy effective action for the Goldstone 
modes along the usual lines.  We introduce the unitary field
$$\Sigma=exp(2i\Phi/f),\eqno(sigma)$$
which transforms as $\Sigma\to U_L\Sigma U_R^\dagger$ with $U_L$ and 
$U_R$ elements of $U(3|3)$.  Because we are dealing here with a graded 
group, in order to build invariants, we need to use the supertrace $str$ 
and the superdeterminant $sdet$ instead of the normal trace and 
determinant, with [\cite{dewitt}]
$$\eqalignno{
str(U)&=tr(A)-tr(B),\cr
sdet(U)&=exp(str\log{(U)})=det(A-CB^{-1}D)/det(B).&(strsdet)
}$$
As one can easily verify, it is this definition of the supertrace that 
respects the cyclic property.
To lowest order in the derivative expansion, and to lowest order in the 
quark masses, the chiral effective lagrangian consistent with our graded 
symmetry group is
$$\cL_0={{f^2}\over 8}str(\partial_\mu\Sigma\partial_\mu\Sigma^\dagger)
-v\;str(\cM\Sigma+\cM\Sigma^\dagger),\eqno(clo)$$
where $\cM$ is the quark mass matrix
$$\cM=\left(\matrix{M&0\cr 0&M\cr}\right),\ \ \ \ \ \ \ \ 
M=\left(\matrix{m_u&0&0\cr 0&m_d&0\cr 
0&0&m_s\cr}\right).\eqno(massmatrix)$$
As before,
$f$ and $v$ are bare coupling constants which are not yet determined at 
this stage.

The symmetry group is broken by the anomaly to the smaller group\break
$[SU(3|3)_L\times SU(3|3)_R]{\bigcirc\kern -0.30cm s\;}U(1)$ (the semidirect
product arises as a consequence of the graded nature of the groups
involved; the details are irrelevant for this talk).  $SU(3|3)$ consists 
of all elements $U\in U(3|3)$ with $sdet(U)=1$.  The anomalous field
is $\Phi_0=(\eta'-\etat')/\sqrt{2}$, where the relative minus sign comes
from the fact that in order to get a nonvanishing triangle diagram, one
needs to choose opposite explicit signs for the quark and ghost quark
loops, due to the different statistics of these fields.  $\eta'$ is the 
field describing the normal $\eta'$ particle, while $\etat'$ is the 
ghost $\eta'$ consisting of ghost quarks and ghost antiquarks.
We will call the field $\Phi_0$ the super-$\eta'$ field.  The field 
$\Phi_0\propto str\log{\Sigma}=\log{sdet\;\Sigma}$ is invariant under 
the smaller symmetry group, and we should include arbitrary functions of 
this field in our effective lagrangian.  In analogy to ref. [\cite{gasleu}], 
the correct chiral effective lagrangian is
$$\eqalignno{
\cL=&V_1(\Phi_0)str(\partial_\mu\Sigma\partial_\mu\Sigma^\dagger)
-V_2(\Phi_0)str(\cM\Sigma+\cM\Sigma^\dagger)\cr
&+V_0(\Phi_0)+V_5(\Phi_0)(\partial_\mu\Phi_0)^2,&(chptlag)
}$$
where the function multiplying $i\;str(\cM\Sigma-\cM\Sigma^\dagger)$ can
be chosen equal to zero after a field redefinition. This lagrangian
describes quenched ChPT systematically, as we will show now with a few
examples.

For our first example, let us isolate just the quadratic terms for 
the fields $\eta'$ and $\etat'$, choosing degenerate quark masses for 
simplicity.  We expand
$$\eqalignno{
V_1(\Phi_0)&={{f^2}\over 8}+\dots,\cr
V_2(\Phi_0)&=v+\dots,\cr
V_0(\Phi_0)&={\rm constant}+\mu^2\Phi_0^2+\dots,&(expand)\cr
V_5(\Phi_0)&=\alpha+\dots,
}$$
and obtain
$$\eqalignno{
\cL(\eta',\etat')=&\half(\partial_\mu\eta')^2
-\half(\partial_\mu\etat')^2+\half\alpha(\partial_\mu\eta'
-\partial_\mu\etat')^2\cr
&+\half m_\pi^2(\eta')^2-\half m_\pi^2(\etat')^2+\half\mu^2(\eta'-\etat')^2
+\dots,&(etaplag)
}$$
where $m_\pi^2=8mv/f^2$. The relative minus signs between the $\eta'$ and
$\etat'$ terms in \eq{etaplag} come from the supertraces in
\eq{chptlag}, and are related to the graded nature of the chiral
symmetry group of quenched QCD.  

The inverse propagator in momentum space,
$$(p^2+m_\pi^2)\left(\matrix{1&0\cr 0&-1}\right)
+(\mu^2+\alpha p^2)\left(\matrix{1&-1\cr -1&1}\right),\eqno(invprop)$$ 
clearly cannot be diagonalized in a $p$ independent way, which is quite
different from what one would expect from a normal field theory!
Treating the $\mu^2+\alpha p^2$ 
term as a twopoint vertex, one can easily show
that the repetition of this vertex on one meson line vanishes, due to
the fact that the propagator matrix $\left(\matrix{1&0\cr 0&-1}\right)$
multiplied on both sides by the vertex matrix $\left(\matrix{1&-1\cr
-1&1}\right)$ gives zero. This result coincides exactly
with what one would expect from the quark flow picture for $\eta'$
propagation, as depicted in fig. 2. The straight-through and double
hairpin contributions do not contain any virtual quark loops, and are
therefore present in the \qa. All other contributions should vanish
because they do contain virtual quark loops, and this is exactly what
happens as a consequence of the (admittedly strange) Feynman rules for
the propagator in the $\eta'$--$\etat'$ sector!  This propagator is 
given by the inverse of \eq{invprop} and reads
$${1\over{p^2+m_\pi^2}}\left(\matrix{1&0\cr 0&-1}\right)
-{{\mu^2+\alpha p^2}\over{(p^2+m_\pi^2)^2}}\left(\matrix{1&1\cr 1&1}\right),
\eqno(prop)$$
in which the two terms correspond to the two first diagrams in fig. 2.
\eject
$\null$
\vskip 3cm\par
\centerline{
Figure 2. {\it Quark flow diagrams for the $\eta'$ propagator in full 
QCD.}}
\par \medskip

From \eq{prop} we learn several things. First, because $\mu^2$, which in
full ChPT would correspond to the singlet part of the $\eta'$ mass,
appears in the numerator, we need to keep the $\eta'$ (and its ghost
partner) in quenched ChPT: it cannot be decoupled by taking $\mu^2$
large. Second, this ``propagator" is definitely sick, due to the double
pole term. It should be stressed here that this double pole term is an
unescapable consequence of quenched QCD, and does not result from our
way of setting up \chpt. In the case of
nondegenerate quark masses, this double pole also shows up in
the $\pi^0$ and $\eta$ propagators, due to mixing with the $\eta'$. I will
return to these strange properties of quenched QCD in section 6.

\vbox{
\vskip 3cm\par
\centerline{
Figure 3. {\it One loop pion selfenergy in quenched ChPT.}}
}
\par \medskip

As a second example, we will consider the (charged) pion selfenergy at 
one loop, again with degenerate quark masses.  I will set $\alpha=0$ for 
simplicity.  At one loop, the pion selfenergy only contains tadpoles, 
with either $\phi$ or $\chi$ lines (\cf\ \eq{octet}) on the loop.  Also, 
on the $\phi$ loop, one can have an arbitrary number of insertions of 
the vertex $\mu^2$ if the internal $\phi$ line is an $SU(3)$ singlet.  
These various contributions are drawn in fig. 3, where a solid line 
denotes a $\phi$ line, a dashed line denotes a $\chi$ line, and a cross
denotes a $\mu^2$ vertex.  One finds that the diagrams with the $\phi$ 
and $\chi$ lines on the loop without any crosses cancel, and then, of 
course, that the diagrams with more than one cross vanish, using our 
earlier result for the $\eta'$--$\etat'$ propagator.  We are left with 
only one term, and the result is
$$\Sigma_\pi(p)={{2m_\pi^2}\over{f^2}}{{\mu^2}\over 3}
\int {{d^4k}\over{(2\pi)^4}}{1\over{(k^2+m_\pi^2)^2}}.\eqno(piself)$$
The pion selfenergy is logarithmically divergent, but this 
nonanalytic term is completely different from those that arise in the 
unquenched theory, as it is proportional to $\mu^2$.  One can easily 
convince oneself that the diagrams in fig. 3 which cancel or vanish 
correspond to diagrams with virtual quark or ghost quark loops in the 
quark flow picture.  (For early discussions of the quenched pion 
selfenergy in the quark flow picture, see refs. [\cite{morel,sharpe1}].)

Before I go on to look at some quantitative results, I would like to
discuss one aspect of the chiral expansion in quenched ChPT. The chiral
expansion is basically an expansion in the pion mass (see \eg\ ref.
[\cite{weinberg}]). However, as we have argued above, in quenched ChPT
there is unavoidably another mass scale, namely the singlet part of the
$\eta'$ mass, $\mu^2$. For our expansion to be systematic as an
expansion in the pion mass, we would have to sum up all orders in
$\mu^2$, at a fixed order in the pion mass. This is clearly a formidable
task. In order to avoid this complication in a systematic way, we can
think of $\mu^2/3$ (which turns out to be the natural parameter as it
appears in the chiral expansion) as an independent small parameter.
To check whether this makes any sense, one may note that from the
experimental value of the $\eta'$ mass one obtains a value
$\mu^2/3\approx (500\;MeV)^2$, which is roughly equal to the kaon mass
squared, $m_K^2$. Of course, for quenched QCD the parameter $\mu^2$ need
not have the same value, after all quenched QCD is a different
theory. A lattice computation of this parameter
[\cite{kuretal}] gives $\mu_{\rm quenched}^2/\mu_{\rm full}^2\approx
0.75$. ($\alpha$ can be estimated from $\eta$--$\eta'$ mixing, and is
very small.) Finally, one may also note that both $\mu^2$ and $\alpha$
are of order $1/N_c$, where $N_c$ is the number of colors [\cite{witven}].
I will return to this point in section 6.

\subhead{\bf 5. Quantitative comparison of quenched and full ChPT}

Let us first consider the quenched result for the ratio of the kaon and
pion decay constants $f_K$ and $f_\pi$ [\cite{us1,uslat92}]. I will set
$\alpha=0$ and take $m_u=m_d\equiv m$:
$$\left({{f_K}\over{f_\pi}}\right)^{\rm 1-loop}_{\rm quenched}=
1+{{\mu^2/3}\over{16\pi^2f_\pi^2}}
\left[{{m_K^2}\over{2(m_K^2-m_\pi^2)}}\log{\left({{2m_K^2}
\over{m_\pi^2}}-1\right)}-1\right]+
{{m_K^2-m_\pi^2}\over{16\pi^2f_\pi^2}}{\hat L}.\eqno(fkfpi)$$
$\hat L$ is again a certain combination of $O(p^4)$ couplings of the 
quenched chiral lagrangian, which does not have to be equal to $L$ in 
\eq{fullfkfpi}, since it comes from a different theory (quenched QCD).
(For example, $L$ is cutoff dependent, while $\hat L$ is not.)
Because of this, the result \eq{fkfpi} is not directly comparable to the 
equivalent result for the full theory.  In other words, in order to 
compare quenched and full QCD, we have to consider 
quantities which are independent of the bare 
parameters of the effective action.  (Alternatively, we would need to 
extract the values of the bare parameters from some independent 
measurement or lattice computation, in this case, we would need 
independent determinations of $L$ in both quenched and full
QCD.)  In the full theory, 
$(f_\eta f_\pi^{1/3})/f_K^{4/3}$ is such a quantity [\cite{gasleu}], but 
in the quenched theory this quantity is not well defined, due to the 
double poles which occur in the propagators of neutral mesons.  

We will therefore choose to consider a slightly different theory, in
which sufficiently many charged 
(\ie\ off-diagonal) mesons are present [\cite{uslat92}]. 
This theory is a theory
with two light quarks $m_u=m_d=m$ and two heavy quarks $m_s=m_{s'}=m'$.
This theory contains a $u\overline d$ pion $\pi$, an $s'\overline s$ 
pion $\pi'$ and a $u\overline s$
kaon $K$, with (tree level) mass relation 
$$m_K^2=\half(m_\pi^2+m_{\pi'}^2).\eqno(massrel)$$
One can show that the ratio $f_K/\sqrt{f_\pi f_{\pi'}}$ is independent
of the low energy constants $L$. For the quenched theory we find 
$$\left({{f_K}\over{\sqrt{f_\pi f_{\pi'}}}}\right)^{\rm 1-loop}_{\rm 
quenched}=
1+{{\mu^2/3}\over{16\pi^2f_\pi^2}}
\left[{{m_\pi^2+m_{\pi'}^2}\over{2(m_{\pi'}^2-m_\pi^2)}}
\log{\left({{m_{\pi'}^2}
\over{m_\pi^2}}\right)}-1\right],\eqno(qratio)$$
whereas in the full theory
$$\left({{f_K}\over{\sqrt{f_\pi f_{\pi'}}}}\right)^{\rm 1-loop}_{\rm
full}=1-{1\over{64\pi^2f_\pi^2}}\left[m_\pi^2\log{\left(
{{m_K^2}\over{m_\pi^2}}\right)}+m_{\pi'}^2\log{\left(
{{m_K^2}\over{m_{\pi'}^2}}\right)}\right].\eqno(fratio)$$
Note again that the logarithms in the quenched and unquenched 
expressions are completely different in origin.  

We may now compare these two expressions using ``real world" data, where 
we'll determine the value of the $\pi'$ mass from the mass relation 
\eq{massrel}.  With $m_\pi=140\;MeV$, $m_K=494\;MeV$ and 
$\mu^2/3=0.75\times(500\;MeV)^2$ we find
$$\eqalignno{
\left({{f_K}\over{\sqrt{f_\pi f_{\pi'}}}}\right)^{\rm 1-loop}_{\rm
quenched}&=1.049,\cr
\left({{f_K}\over{\sqrt{f_\pi f_{\pi'}}}}\right)^{\rm 1-loop}_{\rm
full}&=1.023,&(results)
}$$
a difference of $3\%$.  If we choose $\mu^2/3=(500\;MeV)^2$, we find 
a difference of about $4\%$.  This difference is small.  Note however, 
that this is due to the fact that for this particular ratio, ChPT seems 
to work very well, both for the full and the quenched theories.  If one 
only considers the size of the one loop corrections (the numbers behind 
the decimal point), the quenched and full results are very different.
It is also possible, and in fact not unlikely, that part of the 
difference between the full and quenched theory gets ``washed out" by 
the fact that we are considering a ``ratio of ratios".  It follows that 
the relative difference is a lower bound on the difference between the 
quenched and full values of the decay constants.  For another quantity 
for which the difference between quenched and full ChPT has been 
calculated, see ref. [\cite{uslat92}].

Recently, numerical results for quenched $f_K/f_\pi$ have become 
available which are precise enough to make a comparison with \eq{qratio} 
interesting.  These results are shown in fig. 4 [\cite{rajan94}], 
in which the solid line 
depicts a fit of \eq{qratio} to the numerical results.  The quantity $X$ 
is defined as
$$X={{m_\pi^2+m_{\pi'}^2}\over{2(m_{\pi'}^2-m_\pi^2)}}
\log{\left({{m_{\pi'}^2}
\over{m_\pi^2}}\right)}-1,\eqno(X)$$

$$
\epsfbox{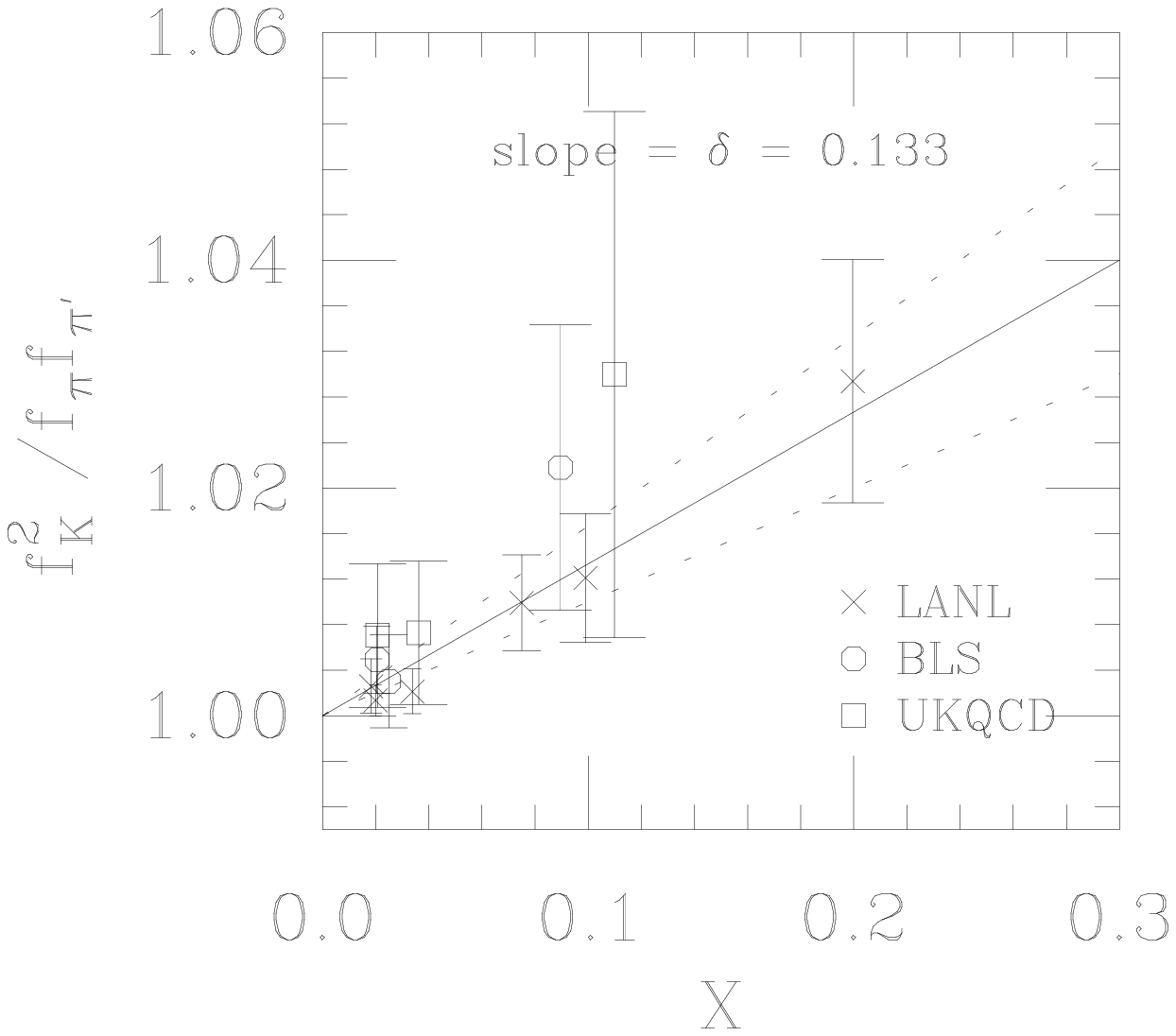}
$$\nobreak
\centerline{
Figure 4. {\it $f_K^2/(f_\pi f_{\pi'})$ versus $X$ (see text for
explanation)
(adapted from ref. [\cite{rajan94}]).}} \medskip

and what was fitted is
$${{f_K^2}\over{f_\pi f_{\pi'}}}=1+\delta X,\eqno(fit)$$
with 
$$\delta={{\mu^3/3}\over{8\pi^2f_\pi^2}}.\eqno(delta)$$
The solid line corresponds to $\delta=0.133$, and the dashed lines 
correspond to a variation of $\pm 0.033$ in this value for $\delta$.  
We expect 
$\delta=0.137$---$0.182$ from \eq{delta}\ for
$\mu^2/3=(0.75$---$1)\times(500\;GeV)^2$.
If 
one plots the same quantity using \eq{fratio}\ instead of \eq{qratio}, 
the results look much less convincing, however, the errors in the 
numerical results are still quite large.  At this stage one can conclude 
that numerical results are consistent with one loop quenched ChPT.  It 
would be interesting to have numerical results with smaller errors.  For 
more detail, see ref. [\cite{rajan94}].

Next, I will review some work on baryons in quenched ChPT 
by Labrenz and Sharpe [\cite{labsha}].  They calculated the one loop 
corrections to the octet baryon mass coming from the cloud of Goldstone 
mesons.  They employed an effective lagrangian for quenched heavy baryon 
ChPT, constructed using the same techniques as described in section 4. 
In the case of degenerate quark masses, the result for the nucleon mass 
is
$$\eqalignno{
m_N=&\mbar-{{3\pi}\over 
2}(D-3F)^2{{\mu^2/3}\over{8\pi^2f_\pi^2}}m_\pi
+2(b_D-3b_F)m_\pi^2\cr
&+\left[{2\over 3}(D-3F)(2D+3\gamma)+{5\over 6}\alpha(D-3F)^2\right]
{{m_\pi^3}\over{8\pi f_\pi^2}}.&(nucleon)
}$$
In this equation, $\mbar$, $D$, $F$, $b_D$, $b_F$ and $\gamma$ are bare 
parameters which occur in the baryon effective action.  $\mbar$ is the 
bare ``average" octet mass, $D$ and $F$ are the well known baryon-meson 
couplings, $b_D$ and $b_F$ are low energy constants which arise as a 
consequence of renormalization (see for instance refs. [\cite{manjen,
berkaimei}]).  $\gamma$ is a new coupling which occurs because of the 
unavoidable presence of the super-$\eta'$ in the \qa.  
The term proportional to 
$\mu^2$ comes from a diagram with a cross on the $\phi$ internal line, 
\ie\ an insertion of the $\mu^2$ twopoint vertex.  Note that in this 
case there are 
also one loop corrections not involving $\mu^2$ which survive 
the \qa, in contrast to the pion selfenergy, \eq{piself}, or 
$f_K/f_\pi$, \eq{fkfpi}.  The authors of ref. 
[\cite{labsha}] then calculated the coefficients using full 
QCD values for the various parameters (from ref. [\cite{berkaimei}]).  
With $\alpha=0$ and $\gamma=0$ ($\gamma=0$ is consistent with available
information, which however is limited [\cite{hatsuda}]), 
they obtained
$$m_N=0.97-0.5{\delta\over{0.2}}m_\pi+3.4m_\pi^2-1.5m_\pi^3,
\eqno(bmass)$$
with $\delta$ as in \eq{delta} and $\delta\approx 0.2$ for 
the full theory.

In ref. [\cite{labsha}], \eq{nucleon} was also compared to recent numerical 
results from ref. [\cite{weinetal}].  These data are presented in fig. 5, 
where the scale $a^{-1}=1.63\;GeV$ is set by $f_\pi$ [\cite{labsha}].  
If one calculates the coefficients in \eq{nucleon} by ``fitting" the 
four data points, one finds
$$m_N=0.96-1.0m_\pi+3.6m_\pi^2-2.0m_\pi^3.\eqno(bfit)$$
This is only four data points for four parameters, and the ``fit" is 
very
sensitive to for instance an additional $m_\pi^4$ term.  
Nevertheless, 
from the agreement between \eq{bmass} and \eq{bfit} it appears that it is 
reasonable to apply ChPT to the results of ref. [\cite{weinetal}].
Note that the individual terms in \eq{bmass} are quite large for the two 
higher pion masses in fig. 5 (this is not unlike the case of unquenched 
ChPT).  From fig. 5 it is also clear that $(\mbar/f_\pi)_{\rm 
quenched}\ne (\mbar/f_\pi)_{\rm full}$ because of the term linear in 
\eq{nucleon}, which is absent in full ChPT.  

Labrenz and Sharpe then went on to consider octet mass splittings.
In order to remove 
effects which can be accomodated by a change of scale, they calculated 
the ratios
$$R_{ij}={{m_i}\over{m_j}},\ \ \ i,j=N,\Lambda,\Sigma,\Xi\eqno(R)$$
in quenched ChPT, and compared these with similar ratios obtained from 
ref. [\cite{berkaimei}].  They assumed that all bare parameters in the 
equations for the octet masses (for explicit expressions, see their 
paper) are equal in the full and quenched theory, and then calculated 
the ratios
$$r_{ij}={{R_{ij}^{\rm quenched}}\over{R_{ij}^{\rm full}}}.\eqno(r)$$
With the assumption that the bare parameters of the quenched and full 
theories are equal, $b_D$ and $b_F$ drop out of the ratios, and with 
$\gamma=0$, $\alpha=0$ and $D$ and $F$ equal to their full QCD values, 
they obtain
$$\eqalignno{
r_{\Sigma N}&=1+0.19(\delta/0.2)+0.13=1.31[1.27]\ \ {\rm for}
\ \delta=0.2[0.15],\cr
r_{\Xi N}&=1-0.46(\delta/0.2)+0.43=0.97[1.09]\ \ {\rm for}
\ \delta=0.2[0.15],&(ratios)\cr
r_{\Lambda N}&=1-0.39(\delta/0.2)+0.26=0.87[0.97]\ \ {\rm for}
\ \delta=0.2[0.15].
}$$
(The choice $\delta=0.15$ corresponds roughly to the value reported in ref. 
[\cite{kuretal}].)
\vfill
\eject
$$
\epsfbox{fig3.ps}
$$\nobreak
Figure 5. {\it The nucleon mass from the lattice [\cite{weinetal}] (copied
from ref. [\cite{labsha}]). The curve is from a fit to the form
$m_N=\mbar+am_\pi+bm_\pi^2+cm_\pi^3$.} \medskip
From this, one would conclude that one can expect errors from quenching 
of at least $20\%$ in the octet splittings.  These differences 
between the quenched and full theories cannot be compensated for by a 
change in scale between quenched and full QCD.

At this point I would like to comment on the above mentioned assumption
that was used in order to obtain \eq{ratios}. 
Let us consider in particular the parameters $b_D$ and $b_F$.
They correspond to higher derivative
terms in the baryon-meson effective action, and are needed in order to
absorb the UV divergences which arise at one loop in ChPT. Since the
size of these divergences is in principle different between the
full and quenched theories, one expects that $b_{D,\rm quenched}$ and
$b_{F,\rm quenched}$ can be different from $b_{D,\rm full}$ and
$b_{F,\rm full}$. 
If we want to
proceed without assuming that the quenched and full $b$'s are equal, 
we have to consider ratios of quantities
independent of the parameters $b_D$ and $b_F$. The situation is
essentially the same as in the case of $f_K/f_\pi$. From the available
results [\cite{labsha}], only one ratio independent of $b_D$ and $b_F$
can be formed:
$$Y={{m_\Sigma m_\Lambda^3}\over{m_N^2 m_\Xi^2}}.\eqno(Y)$$
If we expand $Y$ in the Goldstone meson masses using ChPT, $Y-1$
measures the deviation from the Gell-Mann--Okubo formula (\cf\ ref.
[\cite{berkaimei}] for the full theory).  

Setting $m_\pi=0$ keeping only $m_K$ as in ref. [\cite{labsha}], one 
finds
$$\eqalignno{
Y_{\rm quenched}&=1-1.1046{{m_K^3}\over{8\pi \mbar 
f_\pi^2}}\left(D^2-3F^2\right)+1.3333\delta{{\pi 
m_K}\over{\sqrt{2}\mbar}}\left(D^2-3F^2\right),\cr
Y_{\rm full}&=1-0.4125{{m_K^3}\over{8\pi \mbar 
f_\pi^2}}\left(D^2-3F^2\right).&(Yresults)
}$$
(The parameter $\gamma$ drops out of this particular combination and we 
again take $\alpha=0$.)
These quantities still depend on the other bare parameters, $D$, $F$ and 
$\mbar$.  Again, they could be different in the quenched and full 
theories, and I will leave the quenched values as free parameters.
Substituting $m_K=495$ $MeV$, $f_\pi=132$ $MeV$, $D_{\rm full}=0.75$, 
$F_{\rm full}=0.5$ and $\mbar_{\rm full}=1$ $GeV$ [\cite{berkaimei}]
finally gives
$${{Y_{\rm quenched}}\over{Y_{\rm full}}}=
1-0.0214
+\left[0.293{{\delta}\over{0.2}}-0.306\right]
{{\left(D^2-3F^2\right)_{\rm quenched}}\over{\mbar_{\rm quenched}/
1\ GeV}}.\eqno(Yratio)$$
For any reasonable values of $\mbar$ and $\delta$, and for 
$\left(D^2-3F^2\right)_{\rm quenched}$ not too far from its full theory 
value of $-0.1875$, the difference 
between the quenched and full theories as measured by the ratio $Y_{\rm 
quenched}/Y_{\rm full}$ is not more than a few percent.  Of course the 
same comment that applied in the case of $f_K/f_\pi$ applies here, that 
part of the difference may have been
 washed out by taking ``ratios of ratios".  
Summarizing, the conclusion of this analysis seems to be that the error 
from quenching for octet baryon masses is at least a few percent, and 
could be as much as $20\%$.

\subhead{\bf 6. A sickness of quenched QCD}

Let us again consider the quenched result for 
$f_K/f_\pi$, \eq{fkfpi}, as a
function of the quark masses (using treelevel relations between meson
masses and quark masses),
$$\left({{f_K}\over{f_\pi}}\right)^{\rm 1-loop}_{\rm quenched}=
1+{{\mu^3/3}\over{16\pi^2f_\pi^2}}
\left[{{m_u+m_s}\over{2(m_s-m_u)}}\log{{m_s}\over{m_u}}-1\right]
+{\rm {\hat L}-term}.$$
From this expression it is clear that we cannot take $m_u\to 0$ keeping
$m_s$ fixed, or, to put it differently, that if we take both $m_u$ and
$m_s$ to zero keeping the ratio fixed, the limit depends on this ratio,
and is not equal to one! This is quite unlike the case of full ChPT,
where one can take any quark mass to zero uniformly, and deviations
from $SU(3)$ symmetry due
to this quark mass vanish in this limit. Technically, the reason for this
strange behavior is that there is another mass $\mu$, which, as we
argued before, cannot be avoided in quenched ChPT. This mass is related 
to the singlet part of the $\eta'$ mass, and is not a free parameter of 
(quenched) QCD.  Even if we do not consider any Green's functions with 
$\eta'$ external lines, this mass shows up through the double 
pole term in \eq{prop} on internal lines.  Because of the double pole, 
such contributions can lead to new infrared divergences in the $m_\pi\to 
0$ limit.  This problem with the chiral limit of quenched ChPT shows up 
in other quantities, such as meson masses and $\langle\psib\psi\rangle$ 
[\cite{us1,sharpe2}].

A question one might ask is whether this problem is an artifact of
one loop quenched ChPT [\cite{uslat92}]. For instance, if we would sum
all contributions to the $\eta'$ propagator, maybe the double
pole term would become softer in the $p\to 0$ limit, improving the
infrared behavior of diagrams in which the double pole terms appear. Let
us address this question in the chiral limit, $m_a=0$, where the problem
is most severe. In the full theory, we can write the fully dressed
$\eta'$ propagator as
$${{Z(p)}\over{p^2+\Sigma(p)}},\eqno(fulleta)$$
and define $\mu^2_F(p)=\Sigma(p)$, which onshell is the $\eta'$ mass in 
the chiral limit.  Likewise, in the quenched theory we can write the 
dressed $\eta'$, $\etat'$ propagator as
$$Z_Q(p)\left[{1\over{p^2}}\left(\matrix{1&0\cr 0&-1\cr}\right)
-{{\mu^2_Q(p)}\over{(p^2)^2}}\left(\matrix{1&1\cr 1&1\cr}\right)\right],
\eqno(quenchedeta)$$
which defines $\mu^2_Q(p)$.  To leading order in $1/N_c$, these two 
definitions of $\mu^2(p)$ should lead to the same result:
$$\mu^2_Q(p)=\mu^2_F(p)\left(1+O\left({1\over{N_c}}\right)\right).$$
We also believe that $\mu^2_F(p=0)$ is not equal to zero, since we 
expect the $\eta'$ in the full theory to remain a well-behaved, 
massive particle in the chiral limit.  This 
implies, in sofar as we can rely on the large $N_c$ expansion, that 
$\mu^2_Q(0)\ne 0$, and that the double pole in \eq{prop} is a true 
feature of the theory.  (The argument can be repeated at nonzero quark 
masses, which is necessary if the chiral limit of quenched QCD does not 
exist.)

While this argument is not very rigorous, I believe that the foregoing
discussion implies that the chiral limit of quenched QCD really does not
exist.  This belief is furthermore supported by the following remarks:

$\bullet$ Sharpe [\cite{sharpe2}] has summed a class of diagrams in the 
case of degenerate quark masses for a very simple quantity (the pion 
mass), and found a result that is actually more divergent than the one 
loop result.

$\bullet$ With nondegenerate quark masses there are many more diagrams 
that are infrared divergent in the chiral limit, and it is even less 
probable that resummation will improve the situation.

$\bullet$ Any mechanism improving the infrared behavior would have to 
work for each divergent quantity.  One expects that such a mechanism 
would be related to the double pole term in the $\eta'$ propagator, which 
created the problem in the first place.  But this seems unlikely in view 
of the arguments given above.

$\bullet$ The bare quark mass parameter appearing in the chiral 
effective action is not the same as that appearing in the 
(unrenormalized) QCD lagrangian.  But one can argue that the two bare 
quark masses should be analytically related, and the infrared problem is 
not just a problem of quenched ChPT, but of quenched QCD.

\subhead{\bf 7. Conclusion}

The \qa\ leads to an unknown systematic error in all lattice
calculations that employ this approximation. It would of course be very
nice to have a parameter that interpolates between full and quenched
QCD, and in principle the quark masses could play such a role, since one
expects that quenched QCD corresponds to full QCD with very heavy
quarks. One would have to distinguish here between valence and sea quark
masses, and it is the sea quark mass that would play the role of such a
parameter. This distinction can indeed be made by considering so-called
partially quenched theories [\cite{us2}], but no practical scheme to
implement this idea is known.

Quenched QCD can be defined from a euclidean pathintegral as rigorously
as full QCD. In this talk I have explained that euclidean quenched ChPT
can be used as a tool for a systematic investigation of quenched QCD.
Quenched ChPT does not quite accomplish a task equivalent to that of an
interpolating parameter. Since the bare parameters appearing in the
quenched and full chiral effective actions are not the same (as
explained in section 5) one cannot directly compare quantities
calculated in full and in quenched ChPT. However, one can calculate
combinations of physical quantities which do not depend on the bare
parameters, and in that case a direct comparison between quenched and
full QCD is possible, as we demonstrated with an example involving meson
decay constants. This makes it possible to estimate lower bounds on the
differences which come from quenching; these estimates are dependent on
the values of the meson masses, which can be taken to be the (known)
independent parameters of the theory. For realistic values of these
masses, such differences turn out to be of the order of a few percent
for ratios of decay constants and for baryon octet splittings.

The disadvantage of this more conservative approach is that part of the
difference maybe hidden, because these specific combinations of physical
quantities maybe less sensitive to the effects of quenching
than other quantities of interest. This
is particularly clear in the example of baryon octet masses. In this
case, a comparison based on the assumption that the bare parameters of
the full and quenched effective theories are the same, lead to
differences of up to $20\%$ and more.  Of course, it is not known to 
what extend this assumption is valid.

The differences between the quenched and full theories become markedly
larger for decreasing quark masses. This is due to the fact that new
infrared divergences occur in quenched QCD, which do not have a
counterpart in full QCD. These divergences lead to the nonexistence of
the chiral limit for quenched ChPT (as discussed in section 6). The
origin of this phenomenon can be traced to the special role of the
$\eta'$ in the \qa. In the \qa, the $\eta'$ is a Goldstone boson (it
develops massless poles in the chiral limit), but an additional double
pole term arises in its propagator, rendering it a ``sick" particle. For
nondegenerate quark masses this problem is also inherited by the $\pi^0$
and the $\eta$. In section 6 I argued that the nonexistence of the
chiral limit is a fundamental feature of quenched QCD.

In principle therefore, the chiral expansion breaks down for quenched
QCD. For very small quark masses, at fixed values of the singlet part of
the $\eta'$ mass $\mu^2$, the expansion becomes unreliable. In order to
make progress, one may take the expansion to be an expansion in
$\mu^2/3$ (which was shown to be roughly equal to $m_K^2$
phenomenologically), with coefficients which are functions of the quark
mass. These functions 
sometimes show divergent behavior in the chiral limit (\eg\ the
one loop correction to $f_K/f_\pi$). If such divergences occur, the
expansion is only valid for a range of quark masses which are neither
too small, nor too large. It would be interesting to see whether this
point of view can be made solid.  One way to check whether quenched ChPT 
makes any sense, is to compare its predictions with numerical results.  
Results for $f_K/f_\pi$ (section 5) and $B_V$ (section 3) are consistent 
with one loop quenched ChPT.

It is in principle interesting to study any quantity which is being
computed in quenched lattice QCD in ChPT, for those quantities for which
ChPT is applicable (meson masses, decay constants, condensates and the
kaon B parameter have been calculated [\cite{us1, sharpe1,uslat92}]). As
discussed, this includes not only Goldstone meson physics {\it per se},
but also chiral corrections to baryon masses [\cite{labsha}], and for
the same reason, to mesons containing heavy quarks.

Recently, also attempts have been made to compute pion and nucleon
scattering lengths [\cite{sharpe3,kuretal2}] from quenched lattice QCD.
If one tries to calculate the $I=0$ pion scattering amplitude in
quenched ChPT, one actually finds that the imaginary part is divergent
at threshold, even for nonvanishing pion mass [\cite{pionscat}]! Again,
this can be related to double pole terms in the $\eta'$ propagator.
Apparently euclidean quenched correlation
functions in general cannot be analytically continued to Minkowski
space-time. (The euclidean four pion correlation functions are well
defined.) Further work is needed on pion scattering lengths.

\subhead{\bf Acknowledgments} 

First, I would like to thank Claude Bernard for a
pleasant collaboration, and for very many discussions. I also thank
Steve Sharpe, Jim Labrenz, Akira Ukawa, Hari Dass, Rajan Gupta,
Julius Kuti and Don
Weingarten for discussions. I am grateful for the opportunity to present
these lectures given to me by the organizers of the XXXIV Cracow School
of Theoretical Physics, Zakopane, Poland.
This work is supported in
part by the Department of Energy under contract 
\#DOE-2FG02-91ER40628.

\references

\refis{bijsonwise}
J.~Bijnens, H.~Sonoda and M.~Wise, \prl{53} (1985) 2367.

\refis{langpag}
P.~Langacker and H.~Pagels, \prd{8} (1973) 4595.

\refis{rajan94}
R.~Gupta,  to be published
in the proceedings of the International Symposium on Lattice Field
Theory, Bielefeld, Germany, 1994.

\refis{georgi}
H.~Georgi, {\it Weak Interactions and Modern Particle Physics}, 
Benjamin, 1984.

\refis{oldies}
S.~Weinberg, Phys. Rev. 166 (1968) 1568; S.~Coleman, J.~Wess and B.~Zumino,
Phys. Rev. 177 (1969) 2239; C.G.~Callan, S.~Coleman, J.~Wess and 
B.~Zumino, Phys. Rev. 177 (1969) 2247.

\refis{thooft}
G.~'t~Hooft, \prl{37} (1976) 8.

\refis{cbtasi}
C.W.~Bernard, in {\it From Actions to Answers}, proceedings of the 1989
TASI School, eds. T.~DeGrand and D.~Toussaint, World Scientific, 1990.

\refis{sharpe}
S.R.~Sharpe, in {\it The Fermilab Meeting}, 7th meeting of the DPF, eds.
C.H.~Albright, P.H.~Kasper, R.~Raja and J.~Yoh, World Scientific, 1993.

\refis{mackenzie}
P.B.~Mackenzie, in 
AIP conference proceedings 302 
(XVIth International Symposium on Lepton and Photon 
Interactions), eds. P.~Drell and D.~Rubin, 1994; 
A.S.~Kronfeld and P.B.~Mackenzie,
Ann. Rev. Nucl. Part. Phys. {\bf 43} (1993) 793.

\refis{lat93} 
Lattice'93, proceedings of the International Symposium 
on Lattice Field Theory, 
Dallas, Texas, 1993, published as
Nucl. Phys. {\bf B} (Proc.Suppl.) {\bf 34} (1994).

\refis{quenched}
H.~Hamber and G.~Parisi, \prl{47} (1981) 1792;
E.~Marinari, G.~Parisi and C.~Rebbi, \prl{47} (1981)
1795; D.H.~Weingarten, \plb{109} (1982) 57.

\refis{sharpe1}
S.R.~Sharpe, \prd{41} (1990) 3233; 
Nucl. Phys. {\bf B} (Proc.Suppl.) {\bf 17} (1990) 146;
G.~Kilcup \etal, \prl{64} (1990) 25; S.R.~Sharpe, DOE/ER/40614-5, to be
published in {\it Standard Model, Hadron Phenomenology and Weak Decays
on the Lattice}, ed. G.~Martinelli, World Scientific.

\refis{us1}
C.W.~Bernard and M.F.L.~Golterman, \prd{46} (1992) 853; Nucl. Phys. {\bf 
B} (Proc.Suppl.) {\bf 26} (1992) 360.

\refis{sharpe2}
S.R.~Sharpe, \prd{46} (1992) 3146; Nucl. Phys. {\bf B}(Proc.Suppl.) {\bf 
30} (1993) 213.

\refis{uslat92}
C.W.~Bernard and M.F.L.~Golterman, Nucl. Phys. {\bf B}(Proc.Suppl.) {\bf 30} 
(1993) 217.

\refis{hasleu}
J.~Gasser and H.~Leutwyler, \plb{184} (1987) 83,\plb{188} (1987) 477,
and \npb{307} (1988) 763; H.~Leutwyler, Nucl. Phys. {\bf B} (Proc.Suppl.)
{\bf 4} (1988) 248
and \plb{189} (1987) 197; H.~Neuberger, \npb{300} (1988) 180;
P.~Hasenfratz and H.~Leutwyler, \npb{343} (1990) 241.

\refis{labsha}
J.N.~Labrenz and S.R.~Sharpe, 
Nucl. Phys. {\bf B} (Proc.Suppl.) {\bf 34} (1994) 335.

\refis{morel}
A.~Morel, J. Physique {\bf 48} (1987) 111.

\refis{dewitt}
For a description of the
properties of graded groups, see for example, B.~DeWitt,
{\it Supermanifolds}, Cambridge, 1984.

\refis{gasleu}
J.~Gasser and H.~Leutwyler, \npb{250} (1985) 465.

\refis{weinberg}
S.~Weinberg, Physica {\bf 96A} (1979) 327.

\refis{kuretal}
Y.~Kuramashi, M.~Fukugita, H.~Mino, M.~Okawa and A.~Ukawa, 
\prl{72} (1994) 3448;
Nucl. Phys. {\bf B} (Proc.Suppl.) {\bf 34} (1994) 117.

\refis{witven}
E.~Witten, \npb{156} (1979) 269;
G.~Veneziano, \npb{159} (1979) 213.

\refis{hatsuda}
T.~Hatsuda, \npb{329} (1990) 376; S.R.~Sharpe, private communication.

\refis{manjen}
E.~Jenkins and A.~Manohar, \plb{255} (1991) 558; E.~Jenkins, \npb{368} 
(1992) 190.

\refis{berkaimei}
V.~Bernard, N.~Kaiser and U.~Meissner, Z. Phys. {\bf C60} (1993) 111.

\refis{weinetal}
F.~Butler, H.~Chen, J.~Sexton, A.~Vaccarino and D.~Weingarten,
\prl{70} (1993) 2849; Nucl. Phys. {\bf B} (Proc.Suppl.) {\bf 30} (1993) 377.

\refis{us2}
C.W.~Bernard and M.F.L.~Golterman, \prd{49} (1994) 486;
Nucl. Phys. {\bf B} (Proc.Suppl.) {\bf 34} (1994) 331.

\refis{sharpe3}
R.~Gupta, A.~Patel and S.R.~Sharpe, 
\prd{48} (1993) 388.

\refis{kuretal2}
Y.~Kuramashi, M.~Fukugita, H.~Mino, M.~Okawa and A.~Ukawa,
\prl{71} (1993) 2387.

\refis{pionscat}
C.W.~Bernard, M.F.L.~Golterman, J.N.~Labrenz, S.R.~Sharpe and A.~Ukawa, 
Nucl. Phys. {\bf B} (Proc.Suppl.) {\bf 34} (1994) 334.

\endreferences 

\vfill 
\bye